\documentclass{article}

\usepackage{PRIMEarxiv}

\usepackage[utf8]{inputenc} % allow utf-8 input
\usepackage[T1]{fontenc}    % use 8-bit T1 fonts
\usepackage{hyperref}       % hyperlinks
\usepackage{url}            % simple URL typesetting
\usepackage{booktabs}       % professional-quality tables
\usepackage{amsfonts}       % blackboard math symbols
\usepackage{nicefrac}       % compact symbols for 1/2, etc.
\usepackage{microtype}      % microtypography
\usepackage{lipsum}
\usepackage{fancyhdr}       % header
\usepackage{graphicx}       % graphics
\graphicspath{{media/}}     % organize your images and other figures under media/ folder
\usepackage{algorithmicx} % added
\usepackage{algpseudocode}  % added
\usepackage{amsmath} %added
\usepackage[Algorithmus]{algorithm}
\algnewcommand{\IIf}[1]{\State\algorithmicif\ #1\ \algorithmicthen}
\algnewcommand{\EndIIf}{\unskip\ \algorithmicend\ \algorithmicif}
\algnewcommand{\ElseIIf}{\unskip\ \algorithmicelse\ \algorithmicif}
\usepackage{comment}
\usepackage{multirow}
\title{Efficient Classification of Histopathology Images using Highly Imbalanced Data}

\author{
  Mohammad Iqbal Nouyed \\
  West Virginia University \\
  Morgantown WV 26506, USA\\
  \texttt{monouyed@mix.wvu.edu} \\
   \And
  Mary-Anne Hartley \\
  Yale University School of Medicine \\
  New Haven, CT 06510, USA\\
  \texttt{mary-anne.hartley@yale.edu} \\
  \AND
  Gianfranco Doretto, Donald A. Adjeroh\\
  West Virginia University \\
  Morgantown WV 26506, USA\\
  \texttt{\{gianfranco.doretto, donald.adjeroh\}@mail.wvu.edu} \\
}

\begin{document}
\maketitle

\begin{abstract}
This work addresses how to efficiently classify challenging histopathology images, such as gigapixel whole-slide images for cancer diagnostics with image-level annotation. We use images with annotated tumor regions to identify a set of tumor patches and a set of benign patches in a cancerous slide. Due to the variable nature of region of interest the tumor positive regions may refer to an extreme minority of the pixels. This creates an important problem during patch-level classification, where the majority of patches from an image labeled as 'cancerous' are actually tumor-free. This problem is different from semantic segmentation which associates a label to every pixel in an image, because after patch extraction we are only dealing with patch-level labels.Most existing approaches  address the data imbalance issue by mitigating the data shortage in minority classes in order to prevent the model from being dominated by the majority classes. These methods include data re-sampling, loss re-weighting, margin modification, and data augmentation.  In this work, we mitigate the patch-level class imbalance problem by taking a divide-and-conquer approach. First, we partition the data into sub-groups, and define three separate classification sub-problems based on these data partitions.  Then, using an information-theoretic cluster-based sampling of deep image patch features, we sample discriminative patches from the sub-groups. Using these sampled patches, we build corresponding deep models to solve the new classification sub-problems. Finally, we integrate information learned from the respective models to make a final decision on the patches. Our result shows that the proposed approach can perform competitively using a very low percentage of the available patches in a given whole-slide image.
\end{abstract}

% keywords can be removed

\section{Introduction}
\label{sec:intro}

Whole-slide images (WSIs) are a rich source of information in digital histology, where tissue sections are scanned at gigapixel scale at various microscopic magnification levels \cite{zuraw2022whole,kumar2020whole}. However, the size and number of these images pose challenges for machine learning  models. Firstly, the gigapixel resolution creates memory constraints necessitating input fragmentation. Secondly, annotations of the tumor regions may constitute a very tiny portion of the entire WSI which can create a large class imbalance in the training data.  In recent years, deep models like CNNs and Transformer-based weakly supervised learning methods such as multiple-instance learning (MIL) have shown promising results in gigapixel whole slide image classification with varied sizes. In this approach, WSIs are divided into small image tiles or patches and then aggregated in later stages to make prediction using a classifier \cite{NEURIPS2021_10c272d0,Zhang_2022_CVPR,Li_2021_CVPR}.  MIL treats each WSI as a bag containing multiple instances. If any instance of a WSI is disease-positive then the whole bag (WSI) is labeled as disease-positive. An aggregator classifier is used on the instance-level predictions to get the final image level prediction. 
% In last few years, researchers have seen great performance in WSI classification by introducing Transformer module within the aggregation step \cite{VU2023handcraft,zheng2022graph,zheng2023kernel,NEURIPS2021_10c272d0}. 

Real-world datasets often display long-tailed or imbalanced class distributions \cite{FOTOUHI2019imbalanced,wang2019oversampling,deepak2023brain,cong2022imbalanced}. %In this kind of scenario, some classes contain a large number of samples (majority class), and some other have a very limited number of samples (minority class). When we train aggregate classifiers on these imbalanced dataset, the model can become biased towards the majority class and perform poorly on the minority class.
Common approaches to handling data imbalance work by mitigating the data shortage in minority class by data augmentation \cite{Mullick_2019_ICCV,Kim_2020_CVPR,chen2020feature}, margin modification \cite{NEURIPS2019_621461af}, loss re-weighting \cite{Tan_2020_CVPR,pmlr-v80-ren18a,Cui_2019_CVPR}, and data re-sampling \cite{wallace2011imbalance,more2016survey,chawla2002smote,BUDA2018249}. Though these methods have performed well on imbalanced natural image data, they may not be as effective for WSIs. This is because, in the MIL classifier, the WSI is represented as a bag of image tiles of variable sizes \cite{Wang_2021_CVPR}. However, since the area of the image that actually contains tumor in a WSI can be very small, it means that a majority of tiles in an image weakly labeled 'cancerous' actually do not contain tumor, effectively mislabeling (>80\%) of the tiles\cite{Kong_2022_CVPR}. 

To address these challenges, we propose a patch-level classification method that utilizes cluster-based sampling strategy to solve the imbalance problem between tumor and benign class patches and also provides an efficient histopathology image classification framework for resource-constrained scenarios. The main contributions of this work are as follows:
\begin{itemize}
    \item A group based training approach where we divide the data into three specific sets which help us to decompose the original problem into three sub-problems. 
%    and decompose the original train three different binary classification models, 
    Each sub-problem focuses on discriminating between specific binary classification problems and, when combined, solves the original classification challenge effectively. 
    \item A z-score-based stratified sampling on clustered image patches of the three focus data groups, which allows us to sample most of the patch texture variety by selecting patches from all the distance-based intervals from cluster centroid. 
    \item A learning based information integration from the three sub-problems
    %decision fusion of the three models 
    to obtain the final image level predictions.
\end{itemize}

\section{Related Work}

\subsection{Multiple Instance Learning (MIL) for WSI Classification}

A typical MIL method for WSI classification consists of two stages. First, features are extracted from each instance, and then these instance features are aggregated to obtain a bag-level feature. Then, an image (bag) level classifier is trained using the bag-level features and their corresponding labels. Lin et al. \cite{Lin_Zhuang_Yu_Wang_2024} proposed a model-agnostic framework called CIMIL to improve existing MIL models by using a counterfactual inference-based subbag evaluation method and a hierarchical instance searching strategy to help search reliable instances and obtain their accurate pseudo-labels. 
Qu et al. \cite{qu2022dgmil} proposed a feature distribution-guided MIL framework called DGMIL, for both WSI classification and positive patch localization. Shi et al. \cite{shi2020loss} proposed a loss based attention
mechanism, which simultaneously learns instance weights
and predictions, and bag predictions for deep MIL.
Qu et al. \cite{NEURIPS2022_weno} proposed an end-to-end weakly supervised knowledge distillation framework called WENO for WSI classification. %which integrates a bag-level and an instance-level classifier into a knowledge distillation framework to improve classification performance at both levels.  
Li et al. \cite{Li_2021_CVPR} proposed a deep
MIL model, called DSMIL, which jointly learns a patch (instance) and an image (bag) classifier, using a two-stream architecture. Zhang et al. \cite{Zhang_2022_CVPR} proposed to virtually enlarge the number of
bags by introducing the concept of pseudobags, on which
a double-tier MIL framework, called DFTD-MIL, is built to effectively use the
intrinsic feature. Kong et al. \cite{Kong_2022_CVPR} presented an end-to-end CNN model called the Zoom-In network that uses hierarchical attention sampling
to classify gigapixel pathology images with minority-pixel cancer metastases from the CAMELYON16 dataset. Sharma et al. \cite{pmlr-v143-sharma21a} proposed an end-to-end framework named Cluster-to-Conquer (C2C) that clusters the patches from a WSI into k-groups, samples $k'$ patches from each group for training, and uses an adaptive attention mechanism for slide-level prediction. 
% This demonstrated that dividing a WSI into clusters can improve the model training by exposing it to diverse discriminative features extracted from
the patches. Campanella et al. \cite{campanella2019clinical} presented a deep learning system based on multiple instances of learning that uses only the diagnoses reported as labels for training, thereby avoiding expensive and time-consuming pixel-wise manual annotations.  
Lu et al. \cite{lu2021data} reported an interpretable weakly supervised deep-learning method called CLAM that uses attention-based learning to identify sub-regions of high diagnostic value to accurately classify whole slide images. 
Nouyed et al \cite{nouyed2022efficient} addressed the challenge of high resolution image classification using a  discriminative patch selection approach where they embeded their patch selection approach inside a novel classification framework which can support the use of different off-the-shelf deep models. 

While all the works mentioned above focus on solving the problem of patch-level label corruption from weakly assigned labels at the image level, they do not address the frequent issue of patch class imbalance, where the region of interest (ROI) that defines the label occupies a super minority of the image pixel space. Pawlowski et al. \cite{pawlowski2020needles} investigated the performance of CNNs for minority-pixel image classiﬁcation tasks and their results show that by using a training dataset limited in size, 
CNNs fail to generalize well because of the low ROI-to-image ratio. Usually, the object associated with the label occupies a dominant portion of the image. However, in histopathology image classification such as gigapixel whole-slide image classification, there could be datasets where only a very tiny fraction of the image informs the positive label.

\subsection{Long-tailed histopathology image classification}

Long-tailed classification is a well-known research topic in machine learning where the objective is to solve the data imbalance problem \cite{kubat1997addressing,NIPS1998_df12ecd0}. Under-sampling \cite{drummond2003c4,more2016survey,BUDA2018249} and over-sampling \cite{Sarafianos_2018_ECCV,shen2016relay} are common solutions with known trade-offs between bias and accuracy.  %In under-sampling part of the majority class is discarded , and in oversampling repetitive sampling of minority class is applied. 
While over-sampling can lead to overfitting of the minority class \cite{chawla2002smote}, under-sampling has the potential of information loss about the majority class \cite{more2016survey}.  We can also apply data augmentation to amplify the minority classes \cite{Kim_2020_CVPR,chen2020feature}. Another category of data balancing is called loss re-weighting, in which the loss function is modified to increase weight on the minority class samples and decrease weight on the majority class samples \cite{khan2018cost,japkowicz2002class,Cui_2019_CVPR}. But research has shown that loss re-weighting can be ineffective when the datasets are separable \cite{pmlr-v97-byrd19a}.  

\section{Method}
%\begin{comment}
\begin{figure}[!htbp]
  \centering
  \includegraphics[width=\linewidth]{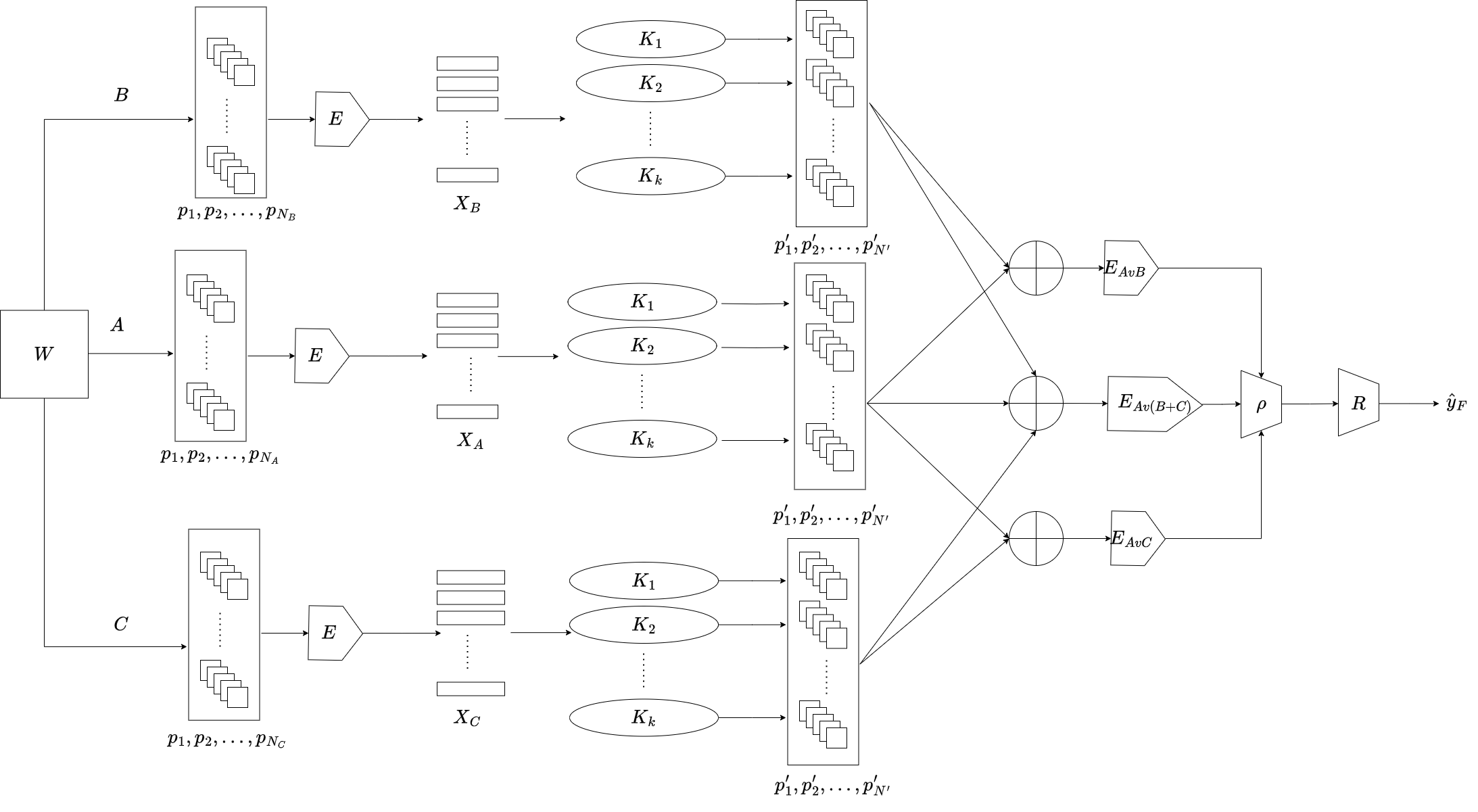}
  \caption{Overview of the proposed framework. At the first stage, all patches of WSIs are extracted  using a pre-trained model  $E$. Then based on available annotation train set data is categorized into 3 data sub-sets. Feature set $X_A, X_B, X_C$ are extracted from each corresponding set. On each set clustering $K$ is performed and then z-score based cluster sampling strategy is applied. Then 3 different models $E_{AvB}, E_{AvC}$ and $E_{Av(B+C)}$ are fine-tuned using the sampled patches $\{p_1', p_2',\dots, p_N'\}$ to train the binary classification models $E_{AvB}, E_{AvC}, E_{Av(B+C)}$. From these, the feature or aggregation information is passed to the aggregation function $\rho(.)$ for patch-level aggregation. And, these aggregated information is used for final patch-level decision fusion using the final $R$ classifier.
  }
  \label{fig:framework}
\end{figure}
\begin{figure}
\includegraphics[width=\textwidth]{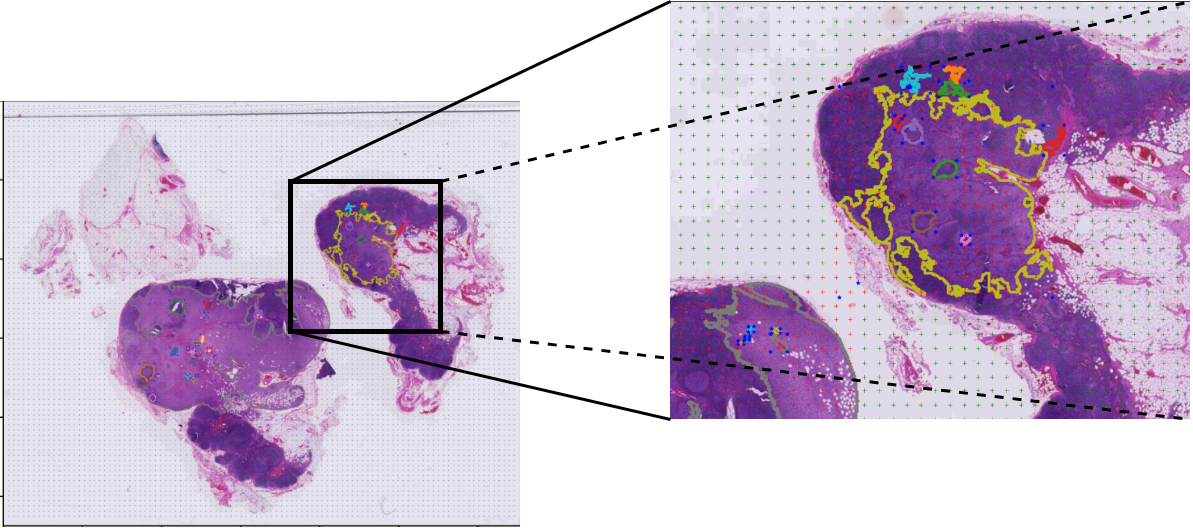}
\caption{Sample WSI, with annotation. Zoomed in section includes annotated regions in different colors, also, the '+' signs indicates the patch $256\times256$ boundaries extracted. } \label{fig:wsi}
\end{figure}
%\end{comment}
In MIL, a group of training samples is considered as a bag containing multiple instances. Each bag has a bag label that is positive if the bag contains at least one positive instance and negative if it contains no positive instance. The instance-level labels are unknown. In the case of binary classification, let $B = \{(x_1, y_1), \dots, (x_n, y_n)\}$ be a bag where $x_i \in \mathcal{X}$ are instances with labels $y_i \in \{0,1\}$, the label of $B$ is given by 

\begin{equation}
    c(B) = 
\begin{cases}
    0,& \text{iff } \sum y_i = 0\\
    1,              & \text{otherwise}
\end{cases}
\end{equation}

First, the image is split into $N \times N$ instances of equal size. We consider the instances from the same image as in the same bag. The main components of our method can be divided into four parts, (1) A divide-and-conquer approach is taken by splitting training data into 3 patch sets using pseudo-labeling and ROI; (2) partitioning of the cancer classification problem into smaller sub-problems based on tumor annotation and source of tissue sample; and (3) Integration of the patch level results using patch level pooling at the feature and prediction levels, followed by activation function and dimensionality reduction (if needed), (4) A threshold percentage of tumor patch per WSI is used to determine the final patch level prediction. Figure \ref{fig:framework} provides an overview of the proposed framework, while Figure \ref{fig:wsi} shows a sample WSI with annotated  tumor region 

\subsection{Partition-based approach to WSI analysis}

Based on the annotation provided in the dataset, we categorize the patches into three different types: 1) Set of tumor patches, denoted as A, so $A$ can be defined as $A = \{ (x, y_p) | y_p = 1, x \in W, y_w=1\}$, where $x$ is an image patch, $y_p$ is patch label, $W$ is an image, and $y_w$ is image label; 2) Set of benign patches that belong to WSIs labeled as cancerous, denoted as set B, so $B = \{ (x, y_p) | y_p = 0, x \in W, y_w=1\}$; (Note that, set B does not indicate a misclassification, a doctor’s misdiagnosis, nor a system’s misdetection. It is simply the set of patches that are extracted outside the annotated tumor regions of the WSI.); and 3) Set of benign patches that belong to WSIs labeled as benign, denoted as set C, so $C = \{ (x, y_p) | y_p = 0, x \in W, y_w=0\}$. The assumption is that the benign patches belonging to a cancerous WSI may contain additional information that can help the model learn to better discriminate between tumor and tumor-free patches. 
%We divide the problem into 
Based on these three data partitions, we can now decompose the original problem into three different binary classification sub-problems: 1) A versus B, 2) A versus C and 3) A versus (B+C). We then train three different classification models of the same architecture for each of the patch-level classification sub-problems. 
% The sampled patches $p_1, p_2, \dots, \p_{N'}$ are passed into three different learning module $G(p_{i=1}^{N'}); \theta: \mathbf{p} \rightarrow \mathbf{z}$ where $\theta$ is the trainable parameters for generating the discriminative representation of the WSI patches between the three patch groups. Using these patch level representation, training is performed using cross-entropy loss. 
% The representation then   is passed through $G: \mathbf{z} \rightarrow \mathbf{y'}$ to obtain slide level prediction probability. 
% All patches from tumor tissue are treated as disease-positive, and all the patches that are coming from benign region of tumor WSI, and all the patches of benign WSIs are treated as disease-negative. 

% We used percentage of predicted tumor patches as threshold to determine WSI-level prediction from patch-level predictions. 
\vspace{-1em}
\subsection{Information theoretic cluster-based sampling}
% \vspace{-3.0em}
\begin{algorithm}[H]
\footnotesize
\caption{Information-theoretic cluster-based patch sampling algorithm}\label{alg:ps_alg}
\begin{algorithmic}[1]
\Require $X, K, |A|$
\Ensure $P$
\For{$K_1, K_2, \dots, K_k$} 
\For{$x \in X$}
\State $d \gets JSD(K_k,x); D(k) \gets D(k) \cup d$ 
%\sqrt{\sum || C_k  - x ||^2} $
\EndFor
\State $D_\sigma(k) \gets \sigma(D(k)); D_\mu(k) \gets \mu(D(k))$
%\State $D_{\mbox{min}}(k) \gets min(D(k))$
%\State $D_{\mbox{max}}(k) \gets max(D(k))$
\For{$d \in D(k)$}
\State $z \gets \frac{d - D_\mu(k)}{D_\sigma(k)}; Z(k) \gets Z(k) \cup z$
%\State $Z_{\mbox{min}}(k) \gets min(Z(k))$
%\State $Z_{\mbox{max}}(k) \gets max(Z(k))$
\EndFor
\EndFor
\For{$K_1, K_2, \dots, K_k$}
\State $S_T = |A| * || D_\mu(k)||$
\For{$x \in X$}
\State $d \gets JSD(K_k,x); z \gets \frac{d - D_\mu(k)}{D_\sigma(k)}; i \gets GetInterval(z); X'(i) \gets x$
\EndFor
\State $s_i = \lfloor S_T / |X'| \rfloor$
\While{$true$} 
\For{$i \in X'$}
\State $\rho \gets RandomSample(X'(i)); P \gets P \cup \rho; S_T \gets S_T - |\rho|$
\EndFor
\State $Update(X'); s_i = \lfloor S_T / |X'| \rfloor$
\If{$S_T \leq 0 \text{ or } s_i \leq 0$} 
\State $break$ 
\EndIf
\EndWhile
\EndFor
\end{algorithmic}
\end{algorithm} 
Because of the partition of the patch sets, we now have a clearer understanding of the class imbalance between the benign and tumor patches. Typically for a dataset $|A| << |B| \text{and } A << |C|$. For this reason, we apply a sampling approach to reduce the class imbalance among set $A$, $B$ and, $C$. First using a pre-trained model, we extract features from all patches. Let $X \gets f(B, \theta)$ or $X \gets f(C, \theta)$ , where $f(., \theta)$ is a feature extractor using the trained parameters $\theta$. We use a parametric clustering method to cluster each of the sets into $k$ different clusters, such that, $K_k$ is the $k$-th cluster centroid. Given a set of patch features $(x_1 , x_2 , \dots, x_n )$, where each patch has been converted to a $d$-dimensional real vector, parametric clustering such as k-means aims to partition the $n$ patches into $k$ clusters $(k \leq n)$ sets $\mathbf{S} = \{S_1, S_2, \dots, S_k \}$ so as to iteratively minimize the within-cluster sum of square errors to reach the local minima or optimum. The objective can be defined as :
\begin{equation}
    \underset{S}{argmin} \sum_{i=1}^k\sum_{\mathbf{x} \in S_i} ||\mathbf{x} - \mathbf{\mu}_i ||^2
\end{equation}
where, $\mathbf{\mu}_i$ is the mean or centroid of the points in $S_i$, $\mathbf{\mu_i} = \frac{1}{|S_i|}\sum_{\mathbf{x}\in S_i}\mathbf{x}$.   We use these $k$ clusters to perform a systematic sampling on the patches such that we can create balanced sets that is not dominated by the minority class.  Algorithm~\ref{alg:ps_alg} shows our procedure for this information-theoretic cluster-based sampling to generating balanced patch sets. Based on the available tumor patches, we sample equal number of patches from each of the $k$ cluster sets, if B' and C' are the new sampled sets such that $B' \subset B$ and $C' \subset C$ then $|B'| = |A|$ , $|C'| = |A|$. Denote $P = B' \mbox{ or } C'$. During clustering we take a stratified random sampling approach based on the Euclidean distance from the cluster centroid in order to maximize the intra cluster variance among the clusters by sampling in such a way that $P$ contains samples from all z-score intervals $X'(i)$. 
For a given patch in a cluster, we represent its computed features as a probability distribution. Similarly with the cluster centroid. We then use an information-theoretic divergence measure, namely the Jensen-Shannon divergence (JSD), to evaluate the dispersion between the patch, and its cluster centroid. 
For two probability distributions $p_1$ and $p_2$, the Jensen-Shannon divergence\cite{cover2006t}  is given by:
\begin{equation}
JSD(p_1,p_2) = \frac{1}{2}D(p_1||q)+\frac{1}{2}D(p_2||q)    
\end{equation}
where $q=\frac{1}{2}(p_1+p_2)$, and $D(p_1||q)$ is the Kullback-Leibler (KL) divergence \cite{cover2006t} between two distributions, given by:
\begin{equation}
    D(p_1||q) = \sum_{c=1}^{|C|} p_1(c)\log \left( \frac{p_1(c)}{q(c)}\right)    
\end{equation}
where $C$ is the number of distinct intervals used in the representation. For each cluster we divide the distribution into intervals based on the z-scores. Then for each patch we calculate the z-score of its dispersion from  the centroid $z \gets \frac{d - D_\mu(k)}{D_\sigma(k)}$. %and, based on that 
Based on this, 
we make sure we uniformly sample from each z-score interval as much as possible so that we can have representation of all possible patch texture variances as much as possible from each cluster, while keeping the total sample size within $S_T$ where $S_T = |A| * || D_\mu(k)||$. Essentially, $S_T$ is the value we get by multiplying the expected total size of the sampled set with normalized mean of all centroid dispersions (or centroid distances). A pseudo-code based description is provided in Algorithm~\ref{alg:ps_alg}.
% This cluster-based approach allows the model to have diverse discriminative patches from WSIs. Instead of local clustering (clustering patches of each WSI) or global clustering (clustering patches of all available WSI) , we take a different clustering approach because the former can too many clusters with repeatable characteristics, where using the latter approach may create clusters based on features that are irrelevant to tumor or benign tissue sample characteristics. Instead we take a semi-clustering approach where we categorize the patches into three distinct groups: tumor patches, benign patches of cancerous WSIs, and benign patches of benign WSIs and train three different patch-level models.  
\subsection{Instance level learning}
The instance-level models encode patches to a $d$-dimensional embedding,\\ $f(\mathbf{x, \theta): \mathbf{x} \rightarrow h}$ where $\theta$ is the set of training parameters. During the training, we use the cross entropy loss on the instance-level labels and prediction of the selected instance to update the classifier's parameters. The loss function for the classifier is define as follows:
\begin{equation}
    L = - \sum_{j} y_j log \hat{y}_j + (1-y_j) log (1-\hat{y}_j),
\end{equation}
 where $y_j$ is the instance-level label. Using the partitioned datasets A, B and C we train three different binary classification models, that learns to discriminate between A vs. B, A vs. C,  and A vs. (B+C), respectively. The objective here is to divide the problem space into sub-problems that discriminate between tumor and benign regions within same tissue image (AvB); between tumor and benign regions of other tissue images (AvC), and tumor and benign regions of both same and other tissue image (Av(B+C). The assumption is that the aggregated feature representations obtained from these expert binary classification models will be more informative for the final prediction. 
 See Figure \ref{fig:framework}.

\subsection{Integrating information from Problem Decompositions}

We investigate information integration from the sub-problems in 
%decision fusion in 
5 different ways: (M0) Majority vote based on the fine-tuned deep model predictions: Let, $\hat{Y}_{AvB} = \{ \hat{y}_i | f_{AvB}^\theta(x_i,y_i) \rightarrow \hat{y}_i \}$, $\hat{Y}_{AvC} = \{ \hat{y}_i | f_{AvC}^\theta(x_i,y_i) \rightarrow \hat{y}_i \}$ and $\hat{Y}_{Av(B+C)} = \{ \hat{y}_i | f_{Av(B+C)}^\theta(x_i,y_i) \rightarrow \hat{y}_i \}$,  are the set of instance-level predictions obtained from models trained on the AvB, AvC, and Av(B+C) datasets. Here $f_{AvB}$, $f_{AvC}$, $f_{Av(B+C)}$ are the binary classification models trained on some parameters $\theta$, and $\hat{y}_i$ is the predicted label of the $i$-th instance. Then we perform a simple majority vote count based on the number of positive predictions to obtain $\hat{y}_F$, the fused label; (M1) Learning-based fusion using Softmax: Let, $S_{AvB} = \{ \sigma_i | \mathcal{S}(f_{AvB}^\theta(x_i,y_i)) \rightarrow \sigma_i \}$, $S_{AvC} = \{ \sigma_i | \mathcal{S}(f_{AvC}^\theta(x_i,y_i)) \rightarrow \sigma_i \}$ and $S_{Av(B+C)} = \{ \sigma_i | \mathcal{S}(f_{Av(B+C)}^\theta(x_i,y_i)) \rightarrow \sigma_i \}$ be a set of instance-level probability distributions, where $\mathcal{S}(.)$ is the softmax function. These instance level probability distributions are concatenated to $x$ and passed to a classifier; (M2) learning-based fusion using feature concatenation followed by dimensionality reduction: Let, $X_{AvB} = \{ x_i | E_{AvB}^\theta(x_i)  \}$, $X_{AvC} = \{ x_i | E_{AvC}^\theta(x_i) \rightarrow x_i \}$ and $X_{Av(B+C)} = \{ x_i | E_{Av(B+C)}^\theta(x_i) \rightarrow x_i \}$, be the set of feature representations obtained from the trained feature encoders $E_{AvB}^\theta, E_{AvC}^\theta, E_{Av(B+C)}^\theta$. These features are concatenated to $x$ and then passed to a dimensionality reduction function $PCA(.)$ followed by a classifier; (M3) Learning-based fusion by applying dimensionality reduction on individual features and then concatenation. Similar to (M2), but here first $PCA(.)$ is applied on individual feature representation and then the reduced features are concatenated; and (M4) Instance level pooling of learned features followed by activation functions and then applying a classifier. Average pooling method on the instance level features is applied to obtain an aggregated patch level representation. This aggregated representation is then passed to GeLU ($G(z)$) function followed by a classifier, where $G(z)$ is defined as follows. 
\begin{equation}
    G(z) = 0.5x(1+tanh[\sqrt{2/\pi}(x+0.044715x^3)])
\end{equation}
Algorithm~\ref{alg:df_alg} shows our proposed procedures for integrating information obtained from solving the three sub-problems. 
%from the out results  the the pseudo-code based description of the decision fusion methods are shown in .

\begin{algorithm}[htbp]
\footnotesize
%\caption{Decision fusion algorithm}
%\label{alg:df_alg}
\caption{Information integration from the problem decomposition}
\begin{algorithmic}[1]
\Require $X_{AvB}, X_{AvC}, X_{Av(B+C)}, \hat{Y}_{AvB}, \hat{Y}_{AvC}, \hat{Y}_{Av(B+C)}, S_{AvB}, S_{AvC}, S_{Av(B+C)}, Y, m$
\Ensure $\hat{y}_F$
\If{$m = 0$}
\For{$\hat{y}_1  \in \hat{Y}_{AvB}, \hat{y}_2  \in \hat{Y}_{AvC}, \hat{y}_3  \in \hat{Y}_{Av(B+C)}$}
\IIf{$\hat{y}_1 + \hat{y}_2 + \hat{y}_3 \geq 2$} 
return $1$ \ElseIIf  return $0$ \EndIIf
\EndFor
\ElsIf{$m = 1$}
\For{$\sigma_1 \in S_{AvB}, \sigma_2 \in S_{AvC}, \sigma_3 \in S_{Av(B+C)}, y \in Y$}
\State $x \gets [\sigma_1; \sigma_2 ; \sigma_3]; \hat{y}_F \gets \text{RandomForest}(x, y); \text{return } \hat{y}_F$
\EndFor
\ElsIf{$m = 2$}
\For{$x_1 \in X_{AvB}, x_2 \in X_{AvC}, x_3 \in X_{Av(B+C)}, y \in Y$}
\State $x \gets [x_1; x_2 ; x_3]; x' \gets PCA(x); \hat{y}_F \gets \text{RandomForest}(x', y); \text{return } \hat{y}_F$
\EndFor
\ElsIf{$m = 3$}
\For{$x_1 \in X_{AvB}, x_2 \in X_{AvC}, x_3 \in X_{Av(B+C)}, y \in Y$}
\State $x'_1 \gets PCA(x_1), x'_2 \gets PCA(x_2); x'_3 \gets PCA(x_3)$
\State $x' \gets [x'_1 ; x'_2 ; x'_3]; \hat{y}_F \gets \text{RandomForest}(x', y); \text{return } \hat{y}_F$
\EndFor
\ElsIf{$m = 4$}
\For{$x_1 \in X_{AvB}, x_2 \in X_{AvC}, x_3 \in X_{Av(B+C)}, y \in Y$}
\State $x'_p \gets AvgPool([x'_1 ; x'_2 ; x'_3]); x' \gets GeLU(x'_p)$
\State $\hat{y}_F \gets \text{RandomForest}(x', y); \text{return } \hat{y}_F$
\EndFor
\EndIf
\end{algorithmic}
\label{alg:df_alg}
\end{algorithm}

\begin{comment}
\begin{figure}[!htbp]
  \centering
  \includegraphics[width=\linewidth]{images/cluster_based_sampling_framework.png}
  \caption{Overview of the proposed framework. At the first stage, all patches of WSIs are extracted  using a pre-trained model  $E$. Then based on available annotation train set data is categorized into 3 data sub-sets. Feature set $X_A, X_B, X_C$ are extracted from each corresponding set. On each set clustering $K$ is performed and then z-score based cluster sampling strategy is applied. Then 3 different models $E_{AvB}, E_{AvC}$ and $E_{Av(B+C)}$ are fine-tuned using the sampled patches $\{p_1', p_2',\dots, p_N'\}$ to train the binary classification models $E_{AvB}, E_{AvC}, E_{Av(B+C)}$. From these, the feature or aggregation information is passed to the aggregation function $\rho(.)$ for patch-level aggregation. And, these aggregated information is used for final patch-level decision fusion using the final $R$ classifier.
  }
  \label{fig:framework}
\end{figure}
\begin{figure}
\includegraphics[width=\textwidth]{images/annotated_wsi.png}
\caption{Sample WSI, with annotation. Zoomed in section includes annotated regions in different colors, also, the '+' signs indicates the patch $256\times256$ boundaries extracted. } \label{fig:wsi}
\end{figure}
\end{comment}
% \vspace{-2em}
\section{Experiments and Results}
\subsection{Database}
We use the publicly available CAMELYON16 dataset for breast cancer metastasis detection.  It has a total of 399 WSIs, with 270 WSIs in training and 129 WSIs in test set. Out of the 270 training images, 111 are tumor WSIs, whose tumor annotation is also provided. For our work, patches of size $256 \times 256$ at 10$\times$ magnification were extracted. Table~\ref{tab:db} provides both slide-level and patch-level database details. Figure~\ref{fig:wsi} shows a sample WSI with tumor annotation.
\begin{table}[!htbp]
  \caption{CAMELYON16 dataset details with info on $256 \times 256$ size patches extracted at $10 \times $ magnification level. 
  }
  \label{tab:db}
  \centering
  \begin{tabular}{|l|l|l|l|}
    \hline
     & & Train & Test\\
    \hline
    \multirow{6}{*}{Image level (WSI)}& Total number  & 270 & 129 \\
      & Positives (number, \%) & 111 (41\%) & 49 (31\%)\\
      & Negatives  (number, \%) & 159 (59\%) & 80 (62\%)\\
    %   & Avg. resolution & (111274, 172237) & (107595,161490) \\
    % & Max resolution &  (221184, 221696) & (212992, 221696) \\
    % & Min resolution & (45056, 35840) & (53248, 28672) \\
      & Avg. area of ROI (pixels, \%) & 444,770 (0.003\%) &  653,670 (0.005\%) \\
    & Max area of ROI (pixels, \%) & 91,418,800 (0.8\%) & 332,954,015 (2.8\%) \\
    & Min area of ROI (pixels, \%) &  10 (0.000\%) &  0 (0.000\%) \\    
    \hline 
    \multirow{4}{*}{Patch level}  & Total number & 4,612,746 & 2,026,538 \\ & Positives (number, \%) & 38,052 (0.8\%)& 31,536 (1.56\%)\\
    & Negatives (number, \%) & 4,574,694 (99\%) & 1,995,131(98\%)\\
    % & Resolution & (256, 256) & (256, 256) \\ 
    \hline 
    \multirow{3}{*}{Patches/Image (PPI)}  & Average & 17083 & 15710 \\
    & Max & 22,787 & 20,906 \\
    & Min  & 1,461 & 3,093 \\
    \hline
  \end{tabular}
\end{table}
\vspace*{-\baselineskip}
\subsection{Architecture and hardware}
For all models, we used ResNet-18 \cite{he2016deep} with a $l=512$ feature representation which was then clustered using k-means with $l2$-normalization. The model was implemented with PyTorch and trained on a single RTX1080 GPU. The models are trained using an SGD optimizer with a batch size of 512 and a learning rate of $1e-4$ for 10 epochs. 
%\section{Training and validation set}
\subsection{Patch labeling}
To partition the patches into groups (A, B, and C), first we find the bounding box around the tumor polygons provided by the CAMELYON16 dataset. After that, for each patch of a WSI, we detect if there is any overlap between the polygon bounding box and the patch coordinates, if there is an overlap we calculate the area of overlapping rectangles using the following formula:
\begin{equation}
(min(x_2, p_2) - max(x_1, p_1)) * (min(y_2, q_2) - max(y_1, q_1))
\end{equation}
where $(x_1,y_1), (x_2,y_2)$ are the polygon bounding box, and $(p_1, q_1), (p_2, q_2)$ are the patch coordinates. We use an overlap threshold to decide whether to assign the patch of a tumor positive WSI in A set, or in the B set, and, if the WSI is tumor negative we put the patches in C set.
\begin{table}[!htbp]
  \caption{Distribution of patches after partitioning into the 3 groups and after applying the clustering based sampling algorithm to create balanced sets. 
  }
  \label{tab:trainval}
  \centering
  \begin{tabular}{|l|l|l|l|l|l|}
    \hline
     & & \textbf{A} & \textbf{B} & \textbf{C} & \textbf{Total}\\
    \hline
    \multirow{3}{*}{Unbalanced data} & Train  & 30,442 & 1,505,762 & 2,153,994 & 3,690,198\\
    & Val  & 7,610 & 376,440 & 538,498 & 922,548 \\ 
    &  Total  & 38,052   &  1,882,202 & 2,692,492 & 4,612,746 \\ 
   \hline
   \multirow{3}{*}{Balanced data} &Train (number, \%) & 30,442 (100\%) & 30,440 (2\%) & 30,440 (1.4\%) & 91,322 (2.4\%) \\
    & Val (number, \%) & 7,610 (100\%)& 7,609 (2\%)& 7,609 (1.4\%)& 22,828 (2.4\%)\\
    & Total (number, \%) & 38,052 (100\%) & 38,049 (2\%)& 38,049 (1.4\%) & 114,150 (2.4\%)\\  \hline 
  \end{tabular}
\end{table}
\subsection{Patch sampling}
We apply a K-means clustering algorithm, with $k=10$, on the pre-trained ResNet18 \cite{he2016deep} features of the unbalanced training set. Now to sample from the B set and C set patches, equal to the size of A set, we use the Euclidean distance from centroid feature to patch feature. The z-score intervals span from -3 to 15, and patches are sampled from within these intervals. Table~\ref{tab:trainval} provides the details on both unbalanced and balanced training and validation datasets. We can observe that for the training and validation set we tried to keep the $A$ set  patches as much as possible so that we don't lose any information regarding tumor presence in the slides. The contributions of the cluster-based sampling strategy or the z-score based sampling strategy were visible, when we compared its performance with just random sampling once the sub-groups of patches are formed. We have found due to the Gaussian nature of the random sampling algorithm most of the patches were similar and does not represent all the variable texture patches within the centroid. This motivated us to take the z-score based sampling approach so that we can properly sample representations from all ranges of variability within a cluster.
\subsection{Efficiency}
From Table~\ref{tab:trainval}, we can observe that we have used 100\% of all A set patches for the training and validation set construction, but reduced the majority classes (namely, class B and class C) down to 2\% and 1.4\% of the original dataset, respectively, in order to match with the minority class. Since during training time these patches are processed sequentially, the time that can be saved can be estimated as $\mathcal{O}\left(B/U\right)$, where $B$ is the total size of the balanced data, and $U$ is the total size of the unbalanced data. Thus, from the table, the proposed method will run about 50 times faster than working without the proposed sampling approach. 
Note that, we are estimating the efficiency gain based on the presence of the balancing step in the proposed framework. %pipeline. 
%Since, other approaches do not take a divide and conquer approach, they do not directly use any under-sampling method that we can compare with. 
We also observe that, speed of convergence is another aspect of measuring the efficiency of the balancing approach which can further establish the efficacy of the balancing step. 

\begin{table}[!htbp]
  \caption{Cross validation result for the 3 models.
  }
  \label{tab:cv}
  \centering
  \begin{tabular}{|l|c|c|c|}
  \hline
  & \textbf{AvB} & \textbf{ AvC} & \textbf{Av(B+C)}\\ \hline 
  & Top-1 Acc. & Top-1 Acc. & Top-1 Acc.\\    
  \hline      
Avg.$\pm$Std. & $0.894 \pm 0.023$ & $ 0.902 \pm	0.022$ & $ 0.897 \pm 0.011$ \\      \hline 
  \end{tabular}
\end{table}
\vspace*{-\baselineskip}
% \vspace{-1em}
\begin{table}[!htbp]
%  \caption{Performance of patch level decsion fusion methods.}
  \caption{Patch-level classification performance using the proposed models for information integrating from the problem decompositions.}
  \label{tab1}
  \label{tab:df}
  \centering
  \begin{tabular}{|l|l|l|l|l|l|}
\hline 
\textbf{Methods} & \textbf{Accuracy} & \textbf{AUC} & \textbf{Precision} & \textbf{Recall}	&  \textbf{F1-score}\\
  \hline
M0 & $0.833 \pm 0.001$& $0.833 \pm 0.001$ & $0.764 \pm 0.001$	& $0.963 \pm 0.001$	& $0.852 \pm0.001$ \\
M1 & $0.980 \pm 0.001$& $0.980 \pm 0.001$ & $0.964 \pm 0.002$ & $0.997 \pm 0.001$ & $0.980 \pm 0.001$ \\
M2 & $0.989 \pm 0.000$	& $0.989 \pm 0.001$ & $0.978 \pm 0.001	$	& $0.999 \pm 0.000$	& $0.988 \pm 0.001$  \\
M3 & $0.988 \pm 0.001$& $0.989 \pm 0.001$ &	$0.978 \pm 0.001$ &	$0.999 \pm 0.000$ &	$0.988 \pm 0.000$ \\
M4 & $0.987 \pm 0.001$& $0.987 \pm 0.004$ & $0.975 \pm 0.001$ &	$0.999 \pm 0.000	$ & $0.987 \pm 0.001$ \\
  \hline 
  \end{tabular}
\end{table}

\begin{table}[!htbp]
  \caption{Comparison of patch level classification performance with the state-of-the-art.
  }
 \begin{minipage}{\linewidth}  
  \label{tab:sota}
  \centering
  \begin{tabular}{|l|l|l|}
\hline 
\textbf{Methods} & \textbf{Accuracy} & \textbf{AUC}\\
  \hline
Loss-ABMIL \cite{shi2020loss} & 0.803 & 0.848\\  
CLAM-SB \cite{lu2021data} & 0.789 &  0.880 \\
CLAM-MB  \cite{lu2021data} & 0.799 & 0.878 \\
DSMIL  \cite{Li_2021_CVPR} & 0.857 & 0.886 \\   
DSMIL+WENO \cite{NEURIPS2022_weno} & 0.901 & 0.930 \\
DTFD-MIL \cite{Zhang_2022_CVPR} & 0.870 & 0.893\\ 
DGMIL \cite{qu2022dgmil} & 0.886 & 0.901 \\
CLAM-SB+CIMIL \cite{Lin_Zhuang_Yu_Wang_2024} & 0.921 & 0.943 \\  
\hline
M2 (Ours) \footnote{\label{tabref}Reported accuracy and AUC is based on validation data.}& $0.989 \pm 0$ & $0.989 \pm 0.001$  \\
M3 (Ours) \footref{tabref}& $0.988 \pm 0.001$ & $0.989 \pm 0.001$  \\
  \hline 
  \end{tabular}
\end{minipage}   
\end{table}

% \vspace{-3.0em}
\subsection{Evaluation}
Using the balanced A, B and C sets, we create 5-fold cross validation sets (80-20 partition). We evaluated performances of the models trained on AvB, AvC and Av(B+C) models, individually and also using feature aggregated decision fusion approaches. 
Table~\ref{tab:cv} shows the individual cross-validation performance of the 3 binary classification models on the balanced datasets. The average top-1 accuracies are $0.894 \pm 0.023$, $0.902 \pm 0.022$ and $0.897 \pm 0.011$ showing strong patch level performance on the individual partitioned data sub-sets. This is using a relatively weak ResNet backbone architecture (ResNet-18).  
Table~\ref{tab:df} shows the performance of different feature aggregation and decision fusion strategies using the combined folds from the 3 partitioned datasets. This makes the folds harder to predict because they include samples from all A, B, and C sets. Even after that we can see that feature concatenation followed by PCA and RF classification ($M2$) shows a strong performance of top-1 accuracy $0.989$ along with high precision ($0.978 \pm 0.001$), recall ($0.999$) and F1-score ($0.988 \pm 0.0005$). The second best method utilizes PCA on deep features following by concatenation of dimensionality reduced features before classification ($M3$), has almost the same performance as $M2$. In fact, except for majority vote approach (M0), all 4 learning-based approaches show strong patch-level classification performance on the validation set. We believe this is 
%an 
indicative of the
%efficiency 
efficacy of our partitioning, sampling, and information integration from the three problem decompositions.  %decision fusion framework. 
Still the work has to show good performance on the test set also, which is much more challenging because we have to infer initial sub-divisions, followed by cluster sampling on an unseen data. 

In Table~\ref{tab:sota} we provide comparative instance-level classification performance results with state-of-the-art methods. For our instance-level classification we used Area Under the Curve(AUC) and Top-1 accuracy as evaluation metrics to compare with other methods. We chose  Loss-ABMIL \cite{shi2020loss}, CLAM-SB \cite{lu2021data},  CLAM-MB  \cite{lu2021data}, DSMIL+WENO \cite{NEURIPS2022_weno}, CLAM-SB+CIMIL \cite{Lin_Zhuang_Yu_Wang_2024}, DSMIL \cite{Li_2021_CVPR}, and DFTD-MIL \cite{Zhang_2022_CVPR}. ABMIL, CLAM, DSMIL models are equipped with specific mechanisms that provide patch prediction, DGMIL is specifically tailored for patch classification. WENO and CIMIL are frameworks for boosting existing MIL models. It can be seen that even with a significantly reduced dataset we were able to achieve the best instance-level performance both in terms of accuracy and AUC. We note that, with very high data imbalance, AUC is a much more effective performance metric than accuracy. However,  since we handled the large class imbalance problem as part of our proposed approach, we believe it is appropriate to then include accuracy for performance measurement.

\section{Conclusion}
In this work, we propose a patch-level classification method that utilizes a group based training approach. By compartmentalizing training into sub-groups, we decompose the original classification problem into smaller classification sub-problems. We then develop models to solve each smaller sub-problem. 
%hypothesize to better handle extreme class imbalances. 
Information from these models are later aggregated using feature and decision fusion approaches leading to a superior classification result. Furthermore, the method also incorporates a cluster-based sampling strategy to solve the significant data imbalance problem between positive and negative classes while maintaining slide-level representation of all WSIs. This allows our approach to efficiently handle a large data source using limited computational resources. Strong patch-level performance in our cross-validation and data fusion experiments validates our claim. Future work should explore the transferable value of the patch level features for slide-level predictions and verify on test set data. 
%Also, we would like to investigate the potential limitations or drawbacks of the proposed approach in greater detail in future work. For instance, 
More ablation studies is needed to further investigate the impact of various algorithmic parameters, e.g, the initial clustering, number of clusters,  complexity of tumor patches, etc.

\vspace{-0.4cm}
\section*{Acknowledgement}
\vspace{-0.3cm}
This work was supported in part by grants from the US National Science Foundation (Award \#1920920 and \#2125872).

\end{document}